\documentclass[preprint,aps,showpacs,,nofootinbib,superscriptaddress,preprintnumbers,epsf,psf]{revtex4}
\usepackage[english]{babel}
\usepackage{pstricks}
\usepackage[dvips]{graphicx}
\usepackage{epsf}
\usepackage{amsmath}
\usepackage{amsfonts}
\usepackage{amssymb}
\usepackage{epsf}
\usepackage[dvips]{graphicx}
\usepackage{dcolumn}% Align table columns on decimal point
\usepackage{bm}% bold math
\everymath{\displaystyle}
\begin{document}
%%%%%%%%%%%%%%%%%%%%%%%%%%%%%%%%%%%%%%%%%%%%%%%%%%%%%%%%%%%%%%%%%%%%
% End of preamble and beginning of text.
%\pagestyle{empty}
%%%%%%%%%%%%%%%%%%%%%%%%%%%%%%%%%%%%%%%%%%%%%%%%%%%%%%%%%%%%%

\title{Velocity-like maximum polarization: irreversibility and quantum measurements}

%\author{\firstname{Oleg}~\surname{Rogachevsky}}
%\email{rogachevsky@jinr.ru} \affiliation{JINR, 141980 Dubna
%(Moscow region), Russia} \affiliation{PNPI RAS, 188300 Gatchina
%(Leningrad district), Russia}
%\author{\firstname{Alexander}~\surname{Sorin}}
%\email{sorin@theor.jinr.ru}
\author{\firstname{Oleg}~\surname{Teryaev}}
\email{teryaev@jinr.ru} \affiliation{Bogoliubov Laboratory of Theoretical Physics \\
and \\
Veksler and Baldin Laboratory of High Energy Physics, \\
Joint Institute for Nuclear Research, 
%\\ and \\
%141980 Dubna
%(Moscow region), Russia} 
%\affiliation{
% Dubna State
%University  \\
141980 Dubna, Russia}

\date{\today}

\begin {abstract}
The polarization emerging in the subsequent scattering processes can never exceed $1$ which corresponds to the fully polarized pure state.
This property is shown to be provided by the addition rule similar to that   
for relativistic velocities never exceeding the speed of light. The cases of spin $1/2$ and $1$ are considered. The photon linear polarization in Thomson scattering is monotonically increasing. This directness is shown to be a consequence of spin measurement procedure  and may be the particular example of ithe anticipated relation between quantum measurement and time irreversibility. The emergent polarization may be considered as a case 
of opposing time's arrows corresponding to microscopic (spin) and macroscopic (momentum) degrees of freedom, respectively.

\end{abstract}


\maketitle

\section{Introduction}

The polarization of initially unpolarized particles in various scattering processes is a rather common phenomenon. It is manifested already in non-relativistic scattering (see e.g. \S 140 in \cite{Landau:1991wop}) in the presence of spin-orbital interaction. The necessary conditions of polarization emergence is the interference of spin-conserving and spin-flip amplitudes with phase difference between them.  

In perturbative QCD this leads to the suppression of the effect \cite{Kane:1978nd} by quark mass and QCD coupling (as the phase can be produced at one-loop order only), and the involvement of various non-perturbative effects including T-odd (i.e., involving non-perturbative QCD phases)  fragmentation functions \cite{Efremov:1992pe}, twist 3 contributions \cite{Efremov:1984ip, Qiu:1991pp}  and Wilson lines 
\cite{Brodsky:2002cx,Collins:2002kn} is required. This puts forward the problem  of total polarization calculation when various mechanisms contribute. Typically the polarization due to each of them is not large (of percent order at best) but what prevents polarization from exceeding one, which is the limit provided by density matrix positivity \cite{Artru:2008cp}, at least in principle?

The similar question is even more sharp in heavy-ion collisions (HIC). 
The polarization in thermodynamic approach \cite{becattini,Becattini:2013fla, becattini2} is proportional to the vorticity\cite{Betz:2007kg} of strongly interacting medium and, formally, may also exceed one. The issue of positivity was addressed \cite{helsep} in the approach \cite{sorin,Sorin:2016smp} to polarization based on axial vortical effect and quark-hadron duality.  The result manifested the compatibilty with density matrix positivity  
and provided the correct estimate of energy dependence and size of polarization prior to the data \cite{STAR:2017ckg}.

The general property of necessity of phase shift in consistent quantum description of global polarization in HIC is realized in the approach \cite{Teryaev:2017wlm,Teryaev:2022pcz} by considering  quantized vortices in pionic superfluid. The baryon polarization is generated in the cores of the vortices and this is accompanied by dissipation being the counterpart of absorptive phases.
Note also that  kinematically vortices provide the link with inclusive polaization \cite{Lisa:2021zkj}.

The scattering may also happen at numerous nucleons, and the suppression of polarization due to the randomization of the orientations of respective scattering planes was suggested \cite{Hoyer:1986pp} as a signal of strongly interacting matter formation. At the same time, the complementary investigation of respective inclusive polarization \cite{Nazarova:2021lry}
provides the quantitative estimates of these randomization effects so that the composition law for polarization generated at different stages of hyperon production process is also of importance.

We start the consideration of composition of polarizations in Section 2 from the case of non-relativistic elastic scattering of spin$-1/2$ beam on spin$-0$ target. We find the complete similarity of composition rule of initial polarization and the one generated in scattering process to the composition rule of {\it velocities} in special relativity. In particular, for addition of non-collinear polarizations the non-commutativity, similar to Thomas precession, is found.

The generalization for higher spin relativistic case is considered in section 3 by the analysis of Compton scattering. We concentrate on photon linear polarization which appears already in Born approximation. The same composition rule is also valid here and leads to the monotonic increase of polarization in subsequent Thomson scattering processes. This directness is appearing because of final state density matrix projection onto that of detector and  may be considered as a particular example of anticipated relation  (see e.g. \S 8 in \cite{Landau:1980mil})
between time irreversibility and quantum measurement.   

The simple generic models for emergent polarization are considered in Section 4. The emergence of polarization is manifesting the decrease of respective entropy  while the entropy corresponding to macroscopic (momentum) degrees of freedom is increasing.  This may be considered as opposing time's arrows with measurement process occuring {\it after} the irreversibility manifestation for microscopic (spin) time's arrow and. corollary,  {\it before} it for the macroscopic (momentum) time's arrow. 

The last section presents main conclusions and outlook of possible further developments. %\cite{Kane:1978nd}

\section{Non-relativistic spin$-1/2$ polarization and relativistic velocity-addition formula}  

Let us start from the non-relativistic scattering amplitude of spin$-1/2$ and spin$-0$ particles \cite{Landau:1991wop} 
\begin{eqnarray}
\label{amp} 
F=a +ib (\vec{\sigma}\vec{n}), 
\end{eqnarray}
where $a$ and $b$ depend on scattering angle $\theta$, $\vec{n}$ is a unit vector normal to the scattering plane and the imaginary unit factor
makes $b$ real in Born approximation. 

Assuming that the initial particles are described by the spin density matrix
\begin{eqnarray}
\label{in} 
\rho_i=\frac{1}{2}(I + (\vec{\sigma}\vec{P})),
 \end{eqnarray}
 where $\vec{P}$ is a polarization vector, final density matrix becomes
 \begin{eqnarray}
\label{fin} 
\rho_f= F \rho_i  F^+ =  \frac{1}{2}(a +ib (\vec{\sigma}\vec{n})(I + (\vec{\sigma}\vec{P}))(a^* - ib^* (\vec{\sigma}\vec{n}).
 \end{eqnarray}
Let us consider first the scattering of unpolarized particls with $P=0$ leading to  
\begin{eqnarray}
\label{unp}   
\rho_f=  \frac{1}{2} (I (|a|^2+|b|^2) + 2 (\vec{\sigma}\vec{n}))\Im (a b^*)), 
\end{eqnarray}
implying that final particles have the polarization 
\begin{eqnarray}
\label{unp}   
\vec P_0 =  \vec{n} \frac{2 \Im (a b^*)}
{|a|^2+|b|^2}.
\end{eqnarray}

This effect may be considered as a manifestation of transition of "macroscopic"  to "microscopic" order. In the initial state the beam has a definite (ordered) momentum  while the spin is completely random. At the same time, in the final state the partial ordering of spin described by $P_0$ is compensated by disorder in momentum described by $a(\theta)$ and $b(\theta)$.  

Let us pass to the case of the partially polarized initial particles, starting from the more simple case when its polarization is  parallel to  
$\vec{n}$, choosing  its direction along the $z$ axis. In that case 
\begin{eqnarray}
\label{finz} 
\rho_f= \frac{1}{2}(a +ib \sigma_z ) (I + \sigma_z P))(a^* - ib^* \sigma_z) = \frac{1}{2} (I (|a|^2+|b|^2+ 2 P \Im (a b^*)) + \sigma_z (P+2\Im (a b^*))).
 \end{eqnarray}
This implies the final polarization which is also, of course, directed along $z$ axis  
\begin{eqnarray}
\label{partz}   
P_f =  \frac{(|a|^2+|b|^2) P+2 \Im (a b^*)}{|a|^2+|b|^2+2 P \Im (a b^*)}=\frac{P+P_0}{1+ P P_0},
\end{eqnarray}
where  (\ref{unp}) is taken into account.

This equation is one of the main results of the current work. The polarization-addition and velocity-addition rules are the same. This relation between properties of non-relativistic spin and relativistic velocity may be due to relation of Lorentz and $SU(2)$ (and $O(3)$) groups. One might recall here the model for Drell-Yan processes\cite{Teryaev:2020tsj} providing the simple geometric description of the result of Feynman diagarams calculation and allowing for the complementary interpretation of experimental data   \cite{Peng:2015spa,Chang:2017kuv}  compatible with general properties of density matrix  \cite{Martens:2017cvj,Gavrilova:2019jea} .   

If one consider polarization as an average  dimensionful  spin and introduce speed of light different from one the relation between the addition rules 
implies also the interchange of fundamental constants 
$$ \hbar \leftrightarrow c. $$ 

Let us now pass to the case of non-collinear $\vec P$ and $\vec n$, than (\ref{fin})  leads to
\begin{eqnarray}
\label{finon} 
\rho_f =  \frac{1}{2} (I (|a|^2+|b|^2   + 2 \Im (a b^*)(\vec P \vec n )) + 2 (\vec{\sigma}\vec{n}) (\Im (a b^*)+|b|^2(\vec P \vec n ) )\nonumber \\+(\vec{\sigma}\vec{P}) (|a|^2-|b|^2) + 2 \Re (a b^*)[\vec P \vec n ] \cdot \vec{\sigma}). 
 \end{eqnarray}
The resulting emergent polarization is
\begin{eqnarray}
\label{partnon}   
\vec{P_f} =\frac{\vec{P_0} + \vec{P}\frac{|a|^2 - |b|^2}{|a|^2 + |b|^2} + \vec{n} \frac{2 |b|^2 (\vec P \vec n ) }{|a|^2 + |b|^2} + [\vec P \vec n ] \frac{2 \Re (a b^*)}
{|a|^2+|b|^2} }{1+ (\vec{P} \vec{P_0})}.
\end{eqnarray}
One can see that it has the components along vectors $\vec n (\parallel \vec P_0), \vec P$ and their vector product. Contrary to one-dimensional case (\ref{partz}) the initial polarization $\vec P$ and the polarization $\vec P_0$, generated in the scattering process enter non-symmetrically. This may be considered as an analog of non-commutativity of Lorentz transformations related to Thomas precession. At the same time, considering the square of polarization, one get after some algebra: 
\begin{eqnarray}
\label{partsq}   
P_f^2 =\frac{(\vec P+\vec P_0)^2-[\vec P \vec P_0]^2}{(1+ (\vec P \vec P_0))^2}.
\end{eqnarray}
This is also completely similar to the addition of non-collinear relativistic velocities.

\section{Polarization-addition in Thomson scattering, irreversibiility and quantum measurement}

We established similarity of addition of {\it relativistic} velocities and polarizations of {\it non-relativistic} spin$-1/2$ particles.
One may ask how the polarizations addition  holds for scattering of relativistic particles with higher spin. 

To address this issue let us consider Thomson scattering. In the case of unpolarized photons it  leads to the emergent linear polarization described bt $\xi_3$ Stokes parameter appearing already in Born approximation. Here we consider what happens if there is also initial photon polarization and 
establish the addition rule. 
   
The phenomenological importance of this effect consists, in paticular, in the manifesrarion of linearly polarized gluons \cite{Pisano:2013cya,Ivanov:2018tvg} as well as tensor polarization of virtual photons and electroweak bosons produced in hadronic \cite{Efremov:1981vs,Peng:2015spa,Chang:2017kuv} and heavy ion \cite{Bratkovskaya:1995kh,Acharya:2019vpe,Buividovich:2012kq,Sheng:2019kmk,Kapusta:2020dco,Goncalves:2021ziy}  collisions. Needless to say, that photon polarization plays also a crucial role in  the search of gravitational waves.

The cross-section with the fixed respective components of polarizations of initial ($\xi_3$)  and final ($\xi^{'}_3$) photons takes the form
 (see e.g. \S 87 in \cite{Berestetskii:1982qgu}):
\begin{eqnarray}
\label{gam} 
d \sigma = \frac{r_e^2}{4} \cdot d \Omega^{'} \cdot \left(\frac{\omega^{'}} {\omega}\right)^2 (F_0+F_{3} (\xi_3+\xi_3^{'})+F_{33}\xi_3\xi_3^{'}),  
 \end{eqnarray}
where $r_e$ is classical electron (or other scatterer) radius,  $\omega(\omega^{'})$ are initial (final)photon frequencies, $d \Omega^{'}=\sin\theta 
d\theta d\phi$ is the solid angle describing the final photon phase, while coefficients $F$ are:
 \begin{eqnarray}
\label{F} 
 F_0 = \frac{\omega} {\omega^{'}}+\frac{\omega^{'}} {\omega} - \sin^2 \theta, \,\
%\nonumber \\
F_{3} = \sin^2 \theta, \,\ 
%\nonumber \\
F_{33} = 1+ \cos^2 \theta.
 \end{eqnarray}  
Let us stress that $\xi_3^{'}$ is the polarization {\it measured} by the detector  which is related (see e.g. \S 65 in \cite{Berestetskii:1982qgu}) to  the polarization $\xi_3^{s}$ emerging in the scattering process by the projection of the respective density matrix ($\rho_f$) onto that of detector ($\rho^{'}$):
\begin{eqnarray}
 \label{meas} 
d \sigma \sim Tr (\rho_s  \rho^{'})  = 1 +\xi_3^s \xi_3^{'}.  
 \end{eqnarray} 
Comparing this with (\ref{gam}) one get 
\begin{eqnarray}
\label{s} 
\xi_3^{s} = \frac{F_{3}+ \xi_3 F_{33}} {F_0+\xi_3 F_{3} }.  
\end{eqnarray}
In particular, for $\xi=0$, the polarization $\xi_s$ emerging in scattering process is 
\begin{eqnarray}
\label{s0} 
\xi_3^{s,0} = 
\frac{F_{3}} {F_0}.  
 \end{eqnarray}
Combining this with (\ref{s}) one get 
\begin{eqnarray}
\label{fs} 
\xi_3^{s} = 
\frac{\xi_3^{s,0}+\frac{F_{33}}{F_0} \xi_3} {1+\xi_3^{s,0} \xi_3}.  
 \end{eqnarray}
As soon as 
\begin{eqnarray}
\label{pos} 
F_{33} \leq {F_0},  
 \end{eqnarray}
which may be in fact related to the general property of  density matrix positivity (see e.g. \cite{Artru:2008cp} and Ref. therein), (\ref{fs}) results in polarization which can never exceed the expression provided by relativistic velocity adding rule. The latter is achieved in the limit when the photon frequency is much smaller than the scatterer (e.g. electron) mass so that the frequency of scattered photon is close to that of initial one 
\begin{eqnarray}
\label{lim} 
\frac{\omega}{m} \to 1, \,\ \omega^{'}  \to \omega,  \,\ F_0 \to  F_{33} = 1+ \cos^2 \theta.
 \end{eqnarray}  
The limiting addition rule
\begin{eqnarray}
\label{ls} 
\xi_3^{s} = 
\frac{\xi_3^{s,0}+\xi_3} {1+\xi_3^{s,0} \xi_3}
 \end{eqnarray}
may be represented  in another form assuming that initial polarization 
$\xi$ also emerged in the scattering process (with corresponding angle
$\theta_0$) and expressing the polarization $\xi_s$ in the similar way in terms of "effective" angle $\theta_s$, so that, together with  (\ref{s0})
in the  limit (\ref{lim}):   
\begin{eqnarray}
\label{an} 
\xi_3 \equiv 
\frac{\sin^2 \theta_0} {1+\cos^2 \theta_0}, \,\ \xi_3^{s0} \equiv 
\frac{\sin^2 \theta_s} {1+\cos^2 \theta}, \,\ \xi_3^{s} \equiv 
\frac{\sin^2 \theta_s} {1+\cos^2 \theta_s}.  
 \end{eqnarray}
As a result, the addition rule (\ref{ls}) takes the simple form   
\begin{eqnarray}
\label{adan} 
\cos^2 \theta_s = \cos^2 \theta_0 \cos^2 \theta_{s0}. 
 \end{eqnarray}
It is interesting to compare this emergent angle with the one appearing
due to scattering corresponding to the "total" angle $\theta_t = \theta_0 +  \theta_{s0}$, 
\begin{eqnarray}
\label{totan} 
\cos^2 \theta_t = (\cos \theta_0 \cos \theta_{s0} - \sin \theta_0 \sin \theta_{s0})^2 . 
 \end{eqnarray}
The cosine (\ref{adan}) corresponding to scattering is larger (and, consequently, the polarization (\ref{an}) is smaller) if the scattering angles  (\ref{totan}) are of the same sign, so that the polarization will be larger after the single scattering described by the angle $\theta_t$. At the same time, if the scattering angles are of different   sign, the polarization due to the single scattering  is smaller than described by (\ref{adan}) and can be even zero if $\theta_0 = -  \theta_{s0}$. 

The consequent scattering processes lead therefore to the monotonically increasing polarization.  Such irreversibility happens despite the conpletely time-reversible dynamics described by QED. This fundamental  property
is reflected in the structure of (\ref{gam}) where the initial and final polarizations  enter completely symmetrically. 

The source of irreversibility is in fact (besides positivity of $F_3$) the quantum measurement process 
(\ref{meas}) leading to expression (\ref{s}). This may be a particular manifestation of the relation of time's arrow and quantum measurement 
anticipated in the classical textbook (see \S 8 in\cite{Landau:1980mil}). This important point deserves a more detailed discussion.

\section{Emergent polarization and opposing time's arrows}

The observed similarity between addition rules of polarization and velocity manifested for both Dirac fermions and photons mark the new observable relation between Lorentz and rotational symmetry. 
The practical applications of the polarization addition rule correspond to the common situations in hadronic in heavy-ion collisions when various dynamical mechanisms contribute to the observable polarization effect. 
For hadronic collisions this may correspond to the addition of T-odd effects in distribution (like Sivers function) and fragmentation (like Collins function) functions. For heavy-ion  collisions the subsequent scattering processes may be considered as a respective counterpart of emergence of global polarization in statistical and anomalous approaches. 
The collective effects here should correspond to the {\it correlation} of orientations of normals to the scattering plane in these processes. 
 
The fundamental aspect of  polarization addition rule is the clear manifestation  of time irreversibility. The most simple case is represented by 
the Thomson scattering of photons. The linear polarization is monotonically increasing in the subsequent scattering processes. 
This irreversiblity appears to be related to the quantum measurement process. The cross section depends on the initial polarization (defined by the experimental setup) and final one (measured by the detector) in the completely symmetric way. The crucial role is played by the measurement process defined as the projection of density matrix corresponding to the {\it real} polarization emerging in the scattering process to the density matrix of the detector. It is this procedure which leads to the velocity-like addition  rule for emergent polarization.  
As soon as the polarizing structure $F_3=\sin^2 \theta$ is always positive,  the polarization is always increasing, which corresponds to the addition of velocities in the same direction. One can recall the role of positivity in QCD evolution equations leading to "Scale's Arrow"( see 4.3.4 in \cite{Artru:2008cp}).

The similar situation holds also for the scattering of Dirac fermions. 
Let us consider the important case of longitudinal polarization 
The generic expression similar to (\ref{gam}) is valid also here, 
\begin{eqnarray}
\label{ferm} 
d \sigma \sim F_0+F_{h} (\xi+\xi^{'})+F_{hh}\xi \xi^{'},  
 \end{eqnarray}
where $\xi, \xi^{'}$ stand  for the helicities in initial and final (i.e., selected by detector) state. The appearance of common factor $F_{h}$ in front of initial $\xi$ and final $\xi^{'}$ helicities is due to time-reversal invariance, while inequality 
\begin{eqnarray}
\label{posith} 
F_0 \geq |F_{hh}|,  
 \end{eqnarray}
completely similar to (\ref{pos}) is providing the analog of (\ref{fs}). This inequality stemming from density matrix positivity is saturated for the {\it chiral} particles leading to the velocity-like addition rule (\ref{partz}). 

Recall, that the irreversibility for photon polarization emerges from the positivity of $F_3$, so that the irreversibility for fermions requires the positivity of $F_h$.  
 This may be provided by parity violation which is the possible mechanisms of polarization, in particular for chiral particles.

In the case of parity conservation the polarization normal to the scattering plane may emerge due to interference of amplitudes with a phase shift
(see Section 2) leading to expression 
\begin{eqnarray}
\label{fermz} 
d \sigma \sim F_0+F_{z} (s_z+s_z^{'})+F_{zz} s_z s_z^{'},  
 \end{eqnarray}
 where only this normal $z-$component of polarization is kept,    
One should stress that $F_z$ is not necessary positive so that the irreversibility depends on the scattering kinematics. More exactly,
$F_z$ is changing sign together with the scattering angle (corresponding to the scattering to the "left" or "right") accompanied by the change of  direction of the normal to the scattering plane. 
Therefore, irreversibility requires scattering in the same direction (say, always to the left) and corresponds to the "angular arrow" (emerging also in QCD evolution at low $x$, see 4.3.3.vii \cite{Artru:2008cp}) rather than times' arrow. 
This property is related to the change of the sign of the {\it vector} polarization of spin $1/2$ particle under $P-$ and $T-$ reflections, while the photon linear polarization  (which is a {\it tensor} one) sign is not changed.

The irreversibility  is therefore including two essential ingredients: positivity and quantum measurement. The latter has some paradoxical specifics: 
the {\it result} of measurement is time-reversible while the state {\it before} measurement, recovered from its result by backward procedure, provides the necessary ingredient of irreversibility. This specifics may be related to the fact that the evolution of spin density matrix manifests the increasing of complexity encoded in polarization. This seeming "violation" of the secomd law of thermodynamic  happens because of decrease of complexity 
related to momentum of scattered particles: while in the initial state the beam consists of particles of definite momentum, the final particles manifest the spread in momentum. One may say that the "macroscopic" (momentum) complexity is transformed to 'microscopic" (spin) one.   

The simplest example of such transformation may be represented by the case, when initial state density matrix corresponds to the absence of "microscopic" spin $J$ polarization (described by Greek indices) and the definite Latin-indexed (discrete) "momentum" state $p$,    
\begin{eqnarray}
\label{in} 
\rho_{\alpha \beta,ij}^{in} = (2J+1)^{-1} \delta_{\alpha \beta} \delta_{i p} \delta_{j p},    
 \end{eqnarray}
 while the final state corresponds to definite spin polarization $\gamma$ and momentum distribution $F_{ij}$: 
\begin{eqnarray}
\label{fin} 
\rho_{\alpha \beta,ij}^{fin} = \delta_{\alpha \gamma} \delta_{\beta \gamma} F_{i j}.    
 \end{eqnarray}
As soon as the density matrix evolutiion is described by the total Hamiltonian $H$,  
\begin{eqnarray}
\label{roev} 
i \frac{ d\rho}{dt} = \{ H, \rho \}, 
 \end{eqnarray}
implying also the same equation for density matrix powers,   
\begin{eqnarray}
\label{ronev} 
i \frac{ d\rho^n}{dt} = \{ H, \rho^n \},
  \end{eqnarray}
one may conclude that 
\begin{eqnarray}
\label{infin} 
Tr (\rho^{in})^n  = Tr (\rho^{fin})^n . 
 \end{eqnarray}
For matrices (\ref{in}, \ref{fin}) this results in
\begin{eqnarray}
\label{infinF} 
Tr (F)^n  = (2J+1)^{1-n} . 
 \end{eqnarray}
This properiy may be realized if $F$ has $2J+1$ non-zero eigenvalues  equal to $(2J+1)^{-1}$ each. One have a sort of interchange between macroscopic and microscopic degrees of freedom, which is similar to  the equality of eigenvalues of density matrices of entangled systems. 
This property can be also seen in
the reduced density matrices, obtained by tracing of macroscopic and microscopic indices, respectively:
\begin{eqnarray}
\label{inr} 
\rho_{\alpha \beta}^{in} = (2J+1)^{-1} \delta_{\alpha \beta},  \\  
 \rho_{ij}^{in} = (2J+1)^{-1} \delta_{i p} \delta_{j p} \\
\rho_{\alpha \beta}^{fin} = \delta_{\alpha \gamma} \delta_{\beta \gamma} \\
\rho_{ij}^{fin} = F_{i j}
\end{eqnarray}
The conditions (\ref{infin}) imply the stability of density matrix eigenvalues and von Neumann entropy \footnote{and Renyi entropy which directly follows from  (\ref{infin}) with $n=2$}.  One is oserving here a sort of entropy flow from microscopic to macroscopic degrees of freedom while the total entropy is conserved. 

This example directly corresponds to the appearance of constant polarization and may be associated, say, with Barnett and Einstein-de Haas effects, or, more close to our consideration, Thomson scattering for $90^0$  to the both left and right leading to the polarization $\xi_3 = 1$.
To adress the more general case of "local" polarization, one may consider the following model of final density matrix, when microscopic 
polarizations $\gamma_k$ are correlated with respective macroscopic distributions $F_k$: 
\begin{eqnarray}
\label{fin1} 
\rho_{\alpha \beta,ij}^{fin} = \sum_k \delta_{\alpha \gamma_k} \delta_{\beta \gamma_k} F^k_{i j}.    
 \end{eqnarray}
The conditions (\ref{infin}) take now the form:
\begin{eqnarray}
\label{infinFk} 
\sum_k Tr (F^k)^n  = (2J+1)^{1-n} . 
 \end{eqnarray}
Let us consider  the simplest case $J=1/2$ when there are only two terms in the sum in (\ref{fin1}) corresponding to the spin projections 
to some axis, $k=\pm$. 
The final state polarization in the state described by the momentum $q$ is  
\begin{eqnarray}
\label{Pfin1} 
P(q) = F^+_{q q} - F^{-} _{q q}.    
 \end{eqnarray}
The non-zero result implies some spin-momentum correlation (in particular, due to spin-otbital interaction, considered above).  

For higher spins there are also other 
components of spin polarization. Say, for $J=1$ there is also a tensor polarization:   
\begin{eqnarray}
\label{Pfin1} 
P_T(q) = F^+_{q q} + F^{-}_{q q} -2 F^{0}_{q q} .    
 \end{eqnarray}

The major problem is the role of quantum measurement in the case of "normal" situation of increasing entropy.
For the case of decreasing entropy and  emergent polarization, the {\it result} of measurement manifests the reversibility, while the irreversibility holds for real emergent polarization {\it before} the measurement. Could it be that for increasing entropy the irreversibility appears after the direct  measurement
of the respective observable? 

At least in the simple models considered above this indeed seems to be the case. The decreasing microscopic entropy is accompanied by the increasing macroscopic one and one may consider that as existence of two "colliding" time's arrows with opposite directions  (c.f.  \cite{Rubino:2020rrj}). 
The irreversibility 
therefore emerges after the measurement according to the microscopic time's arrow and before the measurement according to the macroscopic one.  
Whether such generic considerations may be directly applied in other fields of physics requires further investigations.

\section{Conclusions and Outlook}

The spin quantum measurement (\ref{meas}), together with density  matrix positivity and time reversal {\it invariance}, encoded in  (\ref{gam},\ref{ferm}) plays a crucial role in establishing of the emergent polarization evolutiion similar
to velocity addition rule in special relativity and manifesting some kind of {\it irreversibility}.  

The latter includes the times' arrow manifested in monotonically increasing linear polarization of photons  and angular arrow manifested in the increasing transverse polarization of Dirac fermions. The difference between these two cases is due to positive parity of linear photon polarization and negative parity of Dirac fermion polarization.  

The emergent  polarization may be considered as a transition of order from macroscopic to microscopic degrees of freedom.  This implies opposing time's   arrows for macroscopic and microscopic degrees of freedom.  

The implications of these observations for hadronic collisions correspond to interplay of various mechanisms of generation of various single spin asymmeties, to which the velocity-like rule can be generalized.

For heavy-ion collisions, besides providing the consistent quantum-theoretical treatment of polarization, one may search for opposing time's arrows for spin and hydrodynamical degrees of freedom. Also, the large vorticity and acceleration provides the access to largest ever effective gravity \cite{Kharzeev:2005iz} and emergent conical singularity \cite{Prokhorov:2019yft}. 
It is important, that equivalence principle for spin motion in non-inertial frame is deeply related to 	
quantum measurements \cite{Teryaev:2016edw, Vergeles:2022mqu} and one might expect also the relation to irreversibility in heavy-ion collisions. 

For  more exotic developments, one can look for very different scales. Say, one may thimk whether  cosmological time's arrow can be accompanied by opposing one? It might be also interesting to look for possible interplay of macroscopic and microscopic order in condensed matter systems. One can mention here that, say, the general properties of spin 1 density matrix  \cite{Gavrilova:2019jea} may be related \cite{Goncalves:2021ziy} to qutrit
operation.

\section*{Acknowledgements}

I am indebted to V.I.~Zakharov for stimulating
discussions. 

The work was supported by Russian Science Foundation Grant 21-12-00237.

\end{document}